\documentclass[a4paper,fleqn,usenatbib]{mnras}
\usepackage[caption=false]{subfig}
\usepackage[dvips]{graphicx}
\usepackage{amsmath}
\usepackage{setspace}
\usepackage[T1]{fontenc}
\usepackage{txfonts}

\title[Radio variability and non-thermal components in stars evolving toward PNe]{Radio variability and non-thermal components in stars evolving toward planetary nebulae}
\author[Cerrigone et al.]{L. Cerrigone$^{1, 2}$\thanks{E-mail:
cerrigone@astron.nl}, G. Umana$^{2}$, C. Trigilio$^{2}$\, P. Leto$^{2}$ C. S. Buemi$^{2}$ and A. Ingallinera$^{2}$\\
$^{1}$ASTRON, the Netherlands Institute for Radioastronomy, Dwingeloo, The Netherlands\\
$^{2}$INAF-Catania Astrophysical Observatory, Catania, Italy}
\begin{document}

\date{Accepted 1988 December 15. Received 1988 December 14; in original form 1988 October 11}

\pagerange{\pageref{firstpage}--\pageref{lastpage}} \pubyear{2002}

\maketitle

\label{firstpage}

\begin{abstract}
We present new JVLA multi-frequency measurements of a set of {stars in transition from the post-AGB to the Planetary Nebula phase}  monitored in the radio range over several years. Clear variability is found for five sources. Their light curves show increasing and decreasing patterns.
New radio observations at high angular resolution are also presented for two sources. Among these is IRAS $18062+2410$, whose radio structure is compared to near-infrared images available in the literature. With these new maps, we can estimate inner and outer radii of $0.03''$ and $0.08''$ for the ionised shell, an ionised mass of $3.2\times 10^{-4}$~M$_\odot$, and a density at the inner radius of $7.7\times10^5$~cm$^{-3}$, obtained by modelling the radio shell with the new morphological constraints.

The combination of multi-frequency data and, where available, spectral-index maps leads to the detection of spectral indices  {not due to thermal emission, contrary
to what one would expect in planetary nebulae.} Our results allow us to hypothesise the existence of a link between radio variability and non-thermal emission mechanisms  in the nebulae. This link seems to hold for IRAS $22568+6141$ and may generally hold for those nebulae where the radio flux decreases over time. 
\end{abstract}


\section{Introduction}
After the Asymptotic Giant Branch (AGB), a star evolves on the HR diagram  toward higher temperatures at almost constant luminosity.
What we call a star is actually at this point a system made of a circumstellar nebula and a central hot star whose radiation  makes this nebula glimpse. When the central object is hot enough (T$_\mathrm{eff} \sim 20$--$30\times 10^3$~K), the material in the circumstellar shell starts to be ionised and this event marks the beginning of the planetary nebula (PN) phase.  PNe have been studied for a long time and one of their distinctive traits is an optical spectrum with a great number of nebular emission lines on top of a weak continuum, due to most of the emission from the compact central star being in the ultra-violet region.

The birth of a PN is a qualitatively clear phase of evolution: the star emits more and more photons at shorter wavelengths, which translates in more numerous and stronger emission lines from the nebula as well as a brighter continuum from the growing population of free electrons in it, until the nebula is totally ionised. This free-free continuum is easily detected at radio frequencies, displaying a flat spectrum (optically thin) in the GHz range, with a turning frequency toward optically thick conditions typically between 1 and 5 GHz. The more compact and denser nebulae display a turning frequency close to 5 GHz or higher and the more diluted ones are still optically thin at 1 GHz or below. 

Albeit being clear from a qualitative point of view, the transition from a post-AGB star to a PN is difficult to be investigated because of the paucity of sources in this specific stage and the difficulty in distinguishing these stars from any possible impostor. 

Several authors have pointed their attention toward SAO 244567  (IRAS $17119-5926$, the so-called Stingray Nebula), which has displayed for the last 40 years signs typical of  a central star whose T$_\mathrm{eff}$ is increasing and causing a growing degree of ionisation in its nebula. It seemed then that the Stingray may be  the prototype of these intriguing objects right at the beginning of their PN phase.

 Recent studies by \citet{reindl14} and \citet{schaefer15} have shown that we do not have a clear understanding of the evolutionary status of the Stingray. Its radio flux density is actually decreasing with time \citep{umana08} instead of increasing, as would be expected on the basis of its last 40 yr of evolution \citep{bobrowsky99}.  Several works on similar objects in H$\alpha$ indicate that the onset of ionisation may be accompanied by a variability of unclear origin \citep{arkhipova01_20462}, but the unexpected behaviour of the Stingray consists in a systematic dimming confirmed by a thorough analysis of its photometric variations over decades, which matches with ordinary recombination within the
nebula. This would rather point to a short-duration ionising event in the 1980s \citep{schaefer15}. {Indeed, further studies indicate that this object may have gone through a late thermal pulse and is now moving back toward the AGB \citep{reindl17}.}

To investigate the properties of these short-lived objects in transition from the post-AGB to the PN, we have performed several studies of a sample of hot post-AGB stars in which we detected radio continuum from ionised shells \citep{umana04,umana08,cerrigone08,cerrigone11_radio}. 
To confirm or disregard possible trends of variability detected in our previous works, we will show in this paper new radio data acquired on a sub-sample of our targets where we previously found clear variations or hints for them.

\section{Our sample}
The sample of {sources} in this study has been presented in our previous works, where we first identified stars classified as post-AGB in the literature but with B spectral type, to select objects presumably close to ionising their surroundings. Once successfully detected at 8.4 GHz, we performed multi-frequency and high angular resolution follow-ups, to confirm that the emission arises in circumstellar ionised envelopes. 

These follow-ups pointed our attention to possible variability. We then explicitly targeted this aspect with subsequent observation campaigns. An infrared follow-up was also performed, to study the dust component in the nebulae \citep{cerrigone09}. In Table~\ref{tab:sample}, we list the sources where we have found or suspected radio variability, which are the subject of the present work. For each source, we list the spectral type of the central star {and the evolutionary classification from \citet{suarez06}}. 

\begin{table} \centering
\begin{tabular}{lcc}
\hline
IRAS ID & Sp. Type & Classification \\
\hline
$06556+1623$ &  B:eI/BQ$^a$ & Post-AGB candidate$^a$ \\ 
$17423-1755$ & B7$^a$   & Transition object  \\ 
$17516-2525$ &   em & Post-AGB  \\ 
 $18062+2410$ & B3e &  Transition object  \\ 
$18371-3159$ & B1Iabe$^a$ &  Post-AGB Star (proto-PN)$^a$ \\ 
$18442-1144$ & em$^a$  & Transition object  \\ 
$20462+3416$ & B1.5Ia/Iabe$^a$	  & Transition object  \\ 
$22568+6141$ & B0$^b$ & Planetary Nebula$^a$ \\ 
\hline
\end{tabular}
\begin{flushleft}
\footnotesize{$^a$ SIMBAD classification} \\
\footnotesize{$^b$ \citet{sanchez-contreras12}}  \\
\end{flushleft}
\caption{Sample of nebulae displaying radio variability or suspected to do so, listed with their spectral type and classification from \citet{suarez06}, unless indicated otherwise.}
\label{tab:sample}
\end{table}

\section{Observations and data reduction} \label{sec:observations}
We observed our sources in two different campaigns carried out in 2012 and 2014 with the Karl G. Jansky Very Large Array (JVLA) operated by the National Radio Astronomy Observatory\footnote{The National Radio Astronomy Observatory is a facility of the National Science Foundation operated under cooperative agreement by Associated Universities, Inc.} (NRAO). 

The first campaign was performed between 2012 Jan 27 and 2012 May 24 as project 12A-017, at four frequencies: 1.5, 3.0, 5.0, and 9.0 GHz. The array was in C configuration when sources with declination larger than $-15^\circ$ were observed, while the observations of targets below that value were performed in CnB array. Based on our previous data, we observed IRAS $17423-1755$ at 1.5 GHz in B array, to avoid confusion with a nearby source. The correlator was set up for continuum observations at full polarisation, which gave bandwidths of 1 GHz at the lowest frequency and 2 GHz at the others. The integration time on target ranged from 3 to 10 minutes, depending on the frequency and the expected brightness. Unlike the other targets, IRAS $20462+3416$ was observed at 3.0, 5.0, 9.0, and 45.0 GHz, to investigate the high-frequency range of its spectrum.

In 2014, we observed IRAS $18062+2410$ at high angular resolution at 9.0 and 45 GHz within project 14A-173 on 2014 Feb 23, when the array was in A configuration. As before, the correlator was set up for full-polarisation continuum observations, with band widths of 2 and 8 GHz at 9 and 45 GHz, respectively.

The data were reduced with the Common Astronomical Software Applications (CASA) package, using 3C286 and 3C48 as flux, delay, and bandpass calibrators. First, we searched our data for evident outliers in both frequency and time. Then, we removed shadowed baselines, exact zeroes, \textit{NAN} values, and the first ten seconds of each scan. At the lower frequencies (mostly at 1.5 GHz), substantial interference (RFI) was present. After the first flagging, we then hanning-smoothed the data and performed a run of automated RFI flagging. This produced data sets that needed only minor further editing done manually.

Delay and bandpass calibrations were done on the flux calibrators 3C286 or 3C48, with the exception of  the 45 GHz observations. At the highest frequency,  3C48 was used as a flux calibrator, but it  was  too weak for delay and bandpass calibration, therefore the phase calibrator was used to this aim.

Once delay and bandpass solutions were found, we applied these and calculated amplitude and phase corrections for the flux and complex-gain calibrators. The amplitude of the latter was then scaled on that of the absolute calibrator,  which had been set by applying a model internal to \uppercase{CASA}. When we used the phase calibrator to calibrate  delays and bandpass, a further step was performed in the reduction recipe. To avoid that the spectral slope of the phase calibrator could be transferred to the bandpass solutions, we first bootstrapped it on the absolute calibrator, and then set the values of flux density and spectral index thus derived as model parameters. With these values as a model, we performed again the whole calibration sequence. After a satisfactory calibration was achieved, the amplitude and phase corrections were finally applied to the targets. 

Continuum maps for all targets were created with the \uppercase{clean} task in \uppercase{casa} by applying the multi-frequency synthesis algorithm with some hundreds of iterations (typically between 500 and 1000). 

The flux densities were estimated by fitting a Gaussian to each unresolved source within the \uppercase{casa viewer}  and the rms noise was calculated in an area much larger than the synthesised beam ($>100$ beam), without evident sources in it. {In the attempt to better estimate the spectral indices of the targets, the bandpass at each frequency has been split up in sub-bands, which resulted in slightly different spectral coverage for each source, due to different RFI conditions. Frequencies and flux densities are reported in Table~\ref{tab:results}.}

 IRAS $22568+6141$ was resolved in two lobes, whose flux density was measured by fitting two separate Gaussians. At 45 GHz, IRAS $18062+2410$ was also resolved and its flux density was measured by summing up the emission within a polygon approximately coincident with the $3\sigma$ contour containing the whole source. The results are summarised in Table~\ref{tab:results}, where errors for detected sources are calculated as $\sigma=[rms^2+(0.05\times S_\nu)^2]^{1/2}$, to account for a 5\% indetermination on the absolute calibration.


\begin{table*}
\begin{tabular}{lccclccc}
\hline
Target &  $\nu$  (GHz) &  S$_\nu$ (mJy)  &  $\sigma$ (mJy beam$^{-1}$) & Target &  $\nu$  (GHz) &  S$_\nu$ (mJy)  &  $\sigma$ (mJy beam$^{-1}$) \\
\hline
06556+1623 & 3.02 & 0.23 & 0.05 & 18371-3159 & 1.50 & 0.59 & 0.08 \\ 
 & 3.48 & 0.26 & 0.04 &  & 2.74 & 0.82 & 0.09 \\ 
 & 4.75 & 0.40 & 0.04 &  & 3.56 & 0.88 & 0.08 \\ 
 & 5.25 & 0.37 & 0.04 &  & 4.81 & 0.75 & 0.05 \\ 
 & 5.75 & 0.41 & 0.03 &  & 5.44 & 0.88 & 0.06 \\ 
 & 6.25 & 0.38 & 0.03 &  & 6.00 & 0.72 & 0.05 \\ 
 & 8.24 & 0.48 & 0.04 &  & 8.18 & 0.69 & 0.05 \\ 
 & 8.81 & 0.47 & 0.04 &  & 8.56 & 0.74 & 0.05 \\ 
 &  &  &  &  & 8.94 & 0.58 & 0.06 \\[6pt] 
17423-1755 & 1.50 & -- & 0.05 & 18442-1144 & 1.25 & 11.17 & 0.82 \\ 
 & 2.63 & 0.38 & 0.06 &  & 1.75 & 14.43 & 0.85 \\ 
 & 3.18 & 0.39 & 0.07 &  & 2.63 & 19.17 & 0.96 \\ 
 & 4.68 & 0.41 & 0.05 &  & 3.13 & 21.09 & 1.06 \\ 
 & 5.81 & 0.34 & 0.04 &  & 3.69 & 22.25 & 1.12 \\ 
 & 8.18 & 0.33 & 0.04 &  & 4.74 & 20.77 & 1.04 \\ 
 & 8.56 & 0.30 & 0.04 &  & 5.25 & 21.53 & 1.08 \\ 
 &  &  &  &  & 5.81 & 21.93 & 1.10 \\ 
 &  &  &  &  & 8.24 & 20.67 & 1.03 \\ 
 &  &  &  &  & 8.75 & 20.52 & 1.03 \\[6pt] 
17516-2525 & 1.50 & -- & 0.26 	& 20462+3416 & 2.50 & 0.70 & 0.07 \\ 
 & 2.74 & -- & 0.10 & 	& 2.50 & 0.71 & 0.07 \\ 
 & 3.56 & -- & 0.08 & & 3.50 & 0.66 & 0.06 \\ 
 & 4.75 & 0.25 & 0.04 &  & 3.50 & 0.65 & 0.06 \\ 
  & 5.38 & 0.33 & 0.05 &   & 4.74 & 0.54 & 0.04 \\ 
  & 5.95 & 0.29 & 0.05 &   & 5.25 & 0.43 & 0.04 \\ 
  & 8.18 & 0.42 & 0.04 &   & 5.81 & 0.55 & 0.04 \\ 
  & 8.56 & 0.45 & 0.04 &   & 8.30 & 0.47 & 0.03 \\ 
 & 8.94 & 0.46 & 0.04 &    & 8.93 & 0.47 & 0.03 \\[6pt] 
18062+2410 & 1.50 & -- & 0.14 & 22568+6141 & 1.25 & 17.70 & 1.22 \\ 
 & 2.50 & 0.69 & 0.08 &   & 1.75 & 17.82 & 0.95 \\ 
 & 3.31 & 0.85 & 0.07 &  & 2.50 & 18.16 & 0.91 \\ 
 & 3.69 & 1.25 & 0.09 &   & 3.50 & 17.21 & 0.86 \\ 
 & 4.74 & 1.89 & 0.10 &  & 4.74 & 15.77 & 0.79 \\ 
 & 5.25 & 1.91 & 0.10 &   & 5.25 & 15.95 & 0.80 \\ 
 & 5.81 & 2.15 & 0.12 & & 5.81 & 16.30 & 0.82 \\ 
 & 8.30 & 3.20 & 0.16 & & 8.18 & 13.81 & 0.69 \\ 
 & 8.93 & 3.35 & 0.17 &  & 8.56 & 14.00 & 0.70 \\ 
  & 9.0$^a$ & 3.9 & 0.2 & & 8.94 & 12.40 & 0.62 \\ 
  & 45.0$^a$ & 3.0 & 0.2 & 22568 NW  & 8.18  & 5.9 & 0.3  \\
      & & & & & 8.56 & 6.0 & 0.3  \\
    & & & & & 8.93  & 5.3  & 0.3  \\
    & & & & 22568 SE &  8.18 & 8.2 & 0.4  \\
   & & & &  & 8.56 & 8.0  & 0.4  \\
   & & & &  & 8.93  & 7.4  & 0.4  \\
\hline
{\footnotesize$^a$~Measurement from 2014.}
\end{tabular}
\caption{Flux densities of the targets observed at different frequencies in 2012, except when indicated otherwise. If the target was not detected, the rms noise of the map is listed in mJy~beam$^{-1}$. }
\label{tab:results}
\end{table*}

We also retrieved from the archive of the Australia Telescope National Facility the data at 17 and 19 GHz obtained for the Stingray Nebula within project CX~271 (PI: L. Stavely-Smith) with the Australia Telescope Compact Array (ATCA)  on 2013 Aug 08. The data set was reduced with MIRIAD following the same conceptual recipe as for the other data sets. The results are shown in Table~\ref{tab:stingray}.%

\begin{table} \centering
\begin{tabular}{lcc}
\hline
  Target & \multicolumn{2}{c}{S$_\nu$ (mJy)} \\
   & 17 GHz & 19 GHz \\
  \hline
SAO 244567 &  $23.5\pm0.7$ & $20.7\pm0.6$ \\
\hline
\end{tabular}
\caption{Flux densities of the Stingray nebula from the ATCA project CX271.}
\label{tab:stingray}
\end{table}


{Finally, we also present a map of IRAS~$22568+6141$ obtained with the VLA at 8.4 GHz, when the array was in A configuration on 2006 Feb 3 and 4, with a beam size of about $0.3''$. The inspection of the visibility data indicates that extended flux is missed in these observations. In fact the nominal Largest Angular Scale in A array at 8.4 GHz  is about $5''$, while the source is extended over about $8''$. These observations are then used only to investigate the morphology of the source and not to derive its flux density.}

\section{The continuum spectra} \label{sec:spectra}
\begin{figure*}
  \centering
    \subfloat[\label{06556spec}]{\includegraphics[width=0.3\textwidth]{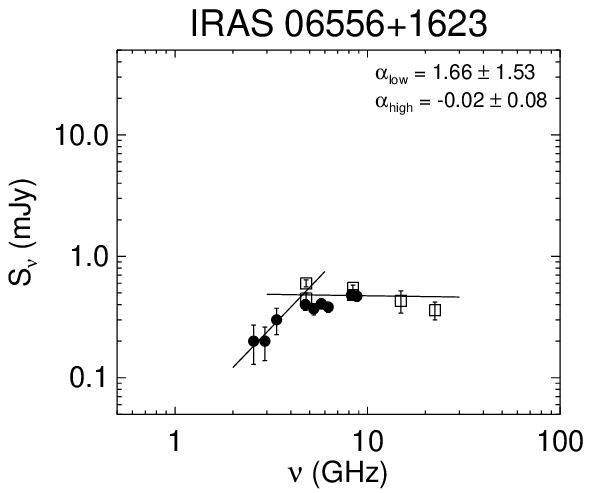}}\qquad
    \subfloat[\label{17423spec}]{\includegraphics[width=0.3\textwidth]{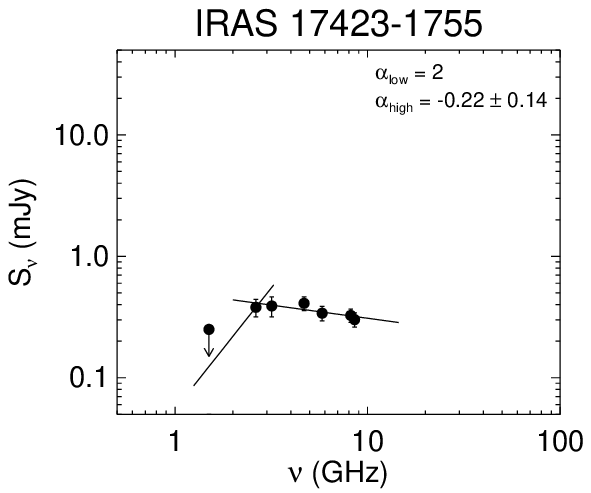}} \qquad
    \subfloat[\label{17516spec}]{\includegraphics[width=0.3\textwidth]{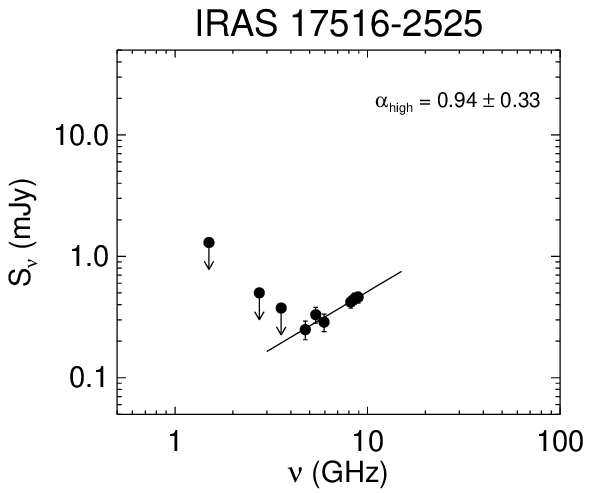}} \\   
     \subfloat[\label{18062spec}]{\includegraphics[width=0.3\textwidth]{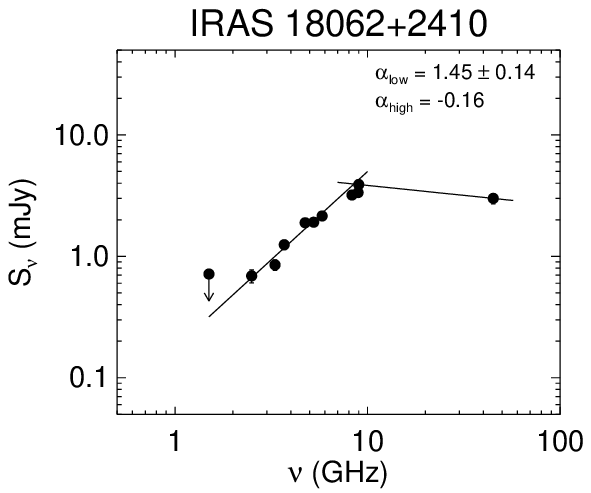}} \qquad 
    \subfloat[\label{18371spec}]{\includegraphics[width=0.3\textwidth]{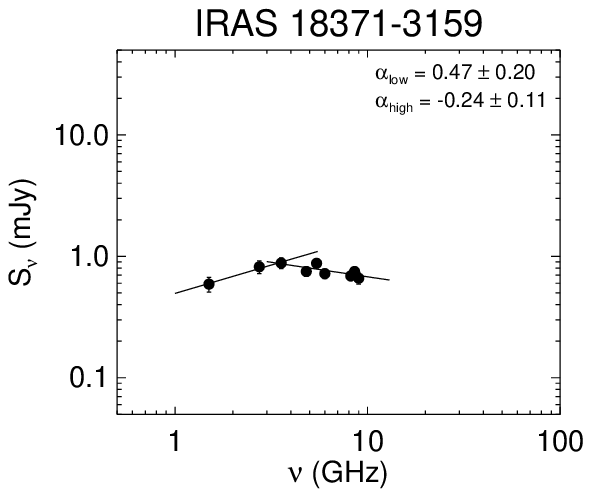}} \qquad
    \subfloat[\label{18442spec}]{\includegraphics[width=0.3\textwidth]{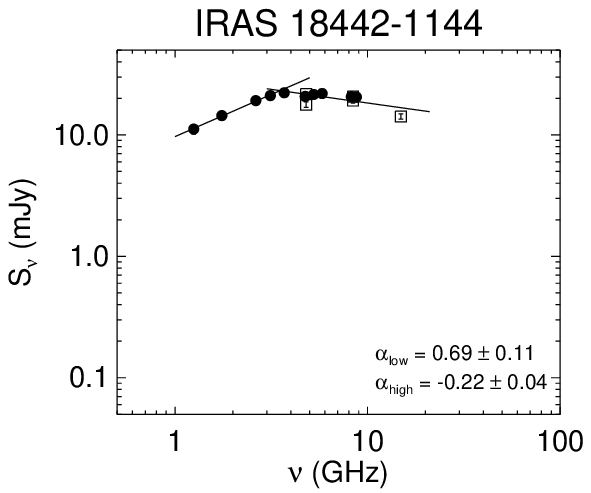}} \\ 
    \subfloat[\label{20462spec}]{\includegraphics[width=0.3\textwidth]{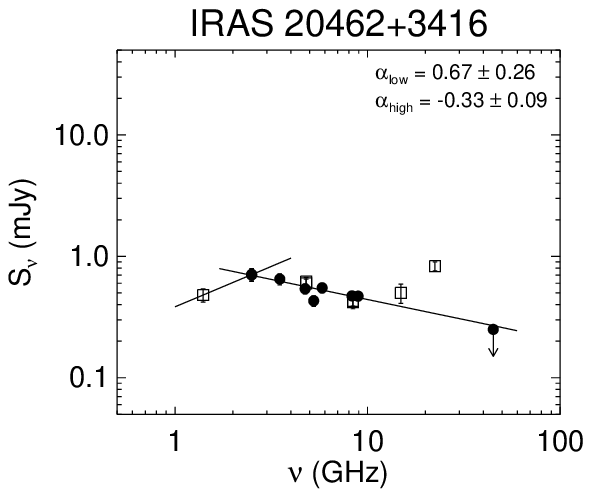}} \qquad 
        \subfloat[\label{22568spec}]{\includegraphics[width=0.3\textwidth]{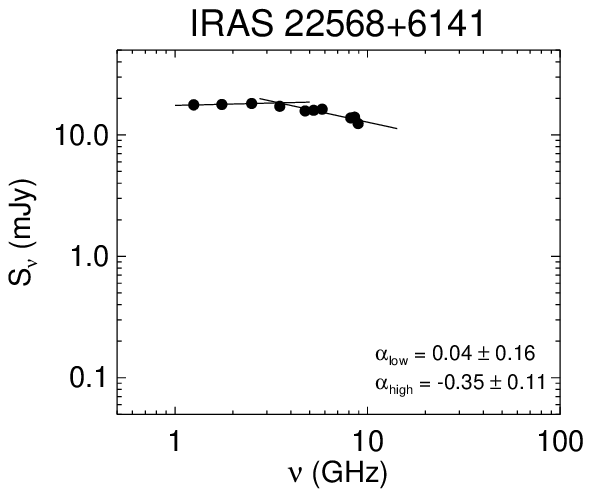}}

    \caption{Continuum spectra of the targets. Data points from this work are solid circles, data from our previous works are open squares.}
    \label{fig:spectra}
  \end{figure*}

Our multi-frequency observations allow us to reconstruct the continuum spectra of our targets (Figure~\ref{fig:spectra}).
The investigation of the continuum spectra of the ionised regions in planetary nebulae is particularly interesting when the evolution can be witnessed over human time scales, because the critical frequency may change with the dilution of the emitting envelope. An example of continuum evolution over decades of observations can be found in \citet{zijlstra08} for NGC 7027. {In Figure~\ref{fig:spectra}, we display the new data from this study and data from our previous works in objects where we conclude that there is no clear hint for variability. These previous points have been taken into account when fitting the emission. Typically, points below 4 GHz have been fitted as optically thick and above 4 GHz as optically thin, with exceptions due to the actual distribution evidently not displaying a turning frequency around 4 GHz. }


We can identify two groups of targets on the basis of their optically-thick slopes. The first group contains IRAS~$06556+1623$, IRAS~$17423-1755$, IRAS~$17516-2525$, IRAS~$18062+2410$, and  {IRAS~$22568+6141$}. {All these sources are close to }what expected for a dense and homogenous emitting region, whose optically-thick spectrum resembles a Planck curve ($S_\nu\sim\nu^2$). 
A spectral index of about 2 {or close to it}  in the radio continuum implies that the ionised region must be close to the central star (where the density is the highest) and small (compact enough to be approximately homogenous). These are conditions expected if the ionisation process is still close to its onset. {In the case of IRAS~$17516-2525$, it is also possible that the emission is due to an ionised stellar wind, since the spectral index is compatible with 0.6 within errors and there is no indication of a turning frequency \citep{panagiafelli}.}

The second group contains sources that deviate {substantially} from the Planck curve, with a spectral index  {around 0.5}. {This group includes IRAS~$18371-3159$, IRAS~$18442-1144$, and IRAS~$20462+3416$. Although the last of these sources was not observed at 1.5 GHz in our JVLA campaign, we can include it in this group thanks to our previous measurements. We measured for IRAS~$20462+3416$ a flux density at 8.4 GHz of $0.42\pm0.05$ mJy in 2003 \citep{cerrigone08}. By comparing this measurement to our new one around the same frequency, we can conclude that we do not see any variability, therefore we can use our previous measurement  of $0.48\pm0.06$ mJy at 1.4 GHz to constrain the spectral index at low frequency of this source. This leads us to a value of about 0.67 between 1.4 and 2.5 GHz.} One may be tempted to sort these sources in an evolutionary sequence depending on how much they deviate from an index of 2, with sources displaying smaller and smaller values of their spectral index during their evolution. Yet, this is not necessarily sensible.  What can be  deduced from a spectral index smaller than 2 is that the ionised region does not match with the conditions that would make its spectrum resemble a Planck curve: it is not very dense and/or not homogenous. These conditions can be achieved by a star whose ionisation process has involved  circumstellar layers that are farther away from the star (i.e., the star has been ionising its nebula for a longer time) or the star is of  lower mass and it started to ionise the nebula later in its life time, when the latter was already diluted.  Another interpretation is that what we are measuring is only the average index over a whole nebula, which can be far from spherical at this point of the stellar evolution, therefore the index cannot be directly linked to the density distribution.

If the emitting region is not homogeneous and displays a radial density distribution, the slope of the thermally thick part of the spectrum can  be linked to the power index of this distribution around the star \citep{pottasch_book}.
The density in the nebula can be expressed as a power $q$  of the radius:
\begin{equation}
\rho (r) = \rho_0 \left[1+\left(\frac{r}{R}\right)^q \right]^{-1}
\end{equation}
where $r$ is the radius of the nebula, $\rho_0$ is the density at the value $R$ of the radius, and $q > {}^3/_2$. It can then be shown that, with such a density profile, the flux density of the source when the opacity is greater than 1 varies as a power of the frequency in the form
\begin{equation}
S_\nu \propto \nu^\frac{2q-3.1}{q-0.5}
\end{equation}

 It should also not be ruled out that the ionisation arises from reasons other than the beginning of the PN phase. Shocks can cause free-free emission and a different evolutionary stage of these sources may  be possible, like the case of the Stingray Nebula. 

 {In the optically-thin regime, all sources display a flat spectrum
within the error bars, except for three of them, namely 
IRAS~$18442-1144$, IRAS~$20462+3416$, and IRAS~$22568+6141$. Surprisingly,
IRAS~$18442-1144$ and IRAS~$20462+3416$ seem to have emission not due to the 
free-free process, as indicated by the steep negative spectral index 
of their continuum spectra.}The presence of non-thermal components in the radio continuum from post-AGB stars  has been established in two such sources, namely IRAS $15103-5754$ and IRAS~$15445-5449$ \citep{perez-sanchez13, suarez15}. This seems to apply also to our two targets. Yet, we believe that, to reach a firm conclusion,  several measurements covering a broad range of frequencies and at different epochs should be obtained, particularly because high-frequency observations are necessary to constrain the optically-thin regime and these may be subject to observational errors due to the atmospheric conditions.

\begin{figure}
  \centering
    {\includegraphics[height=5.5cm]{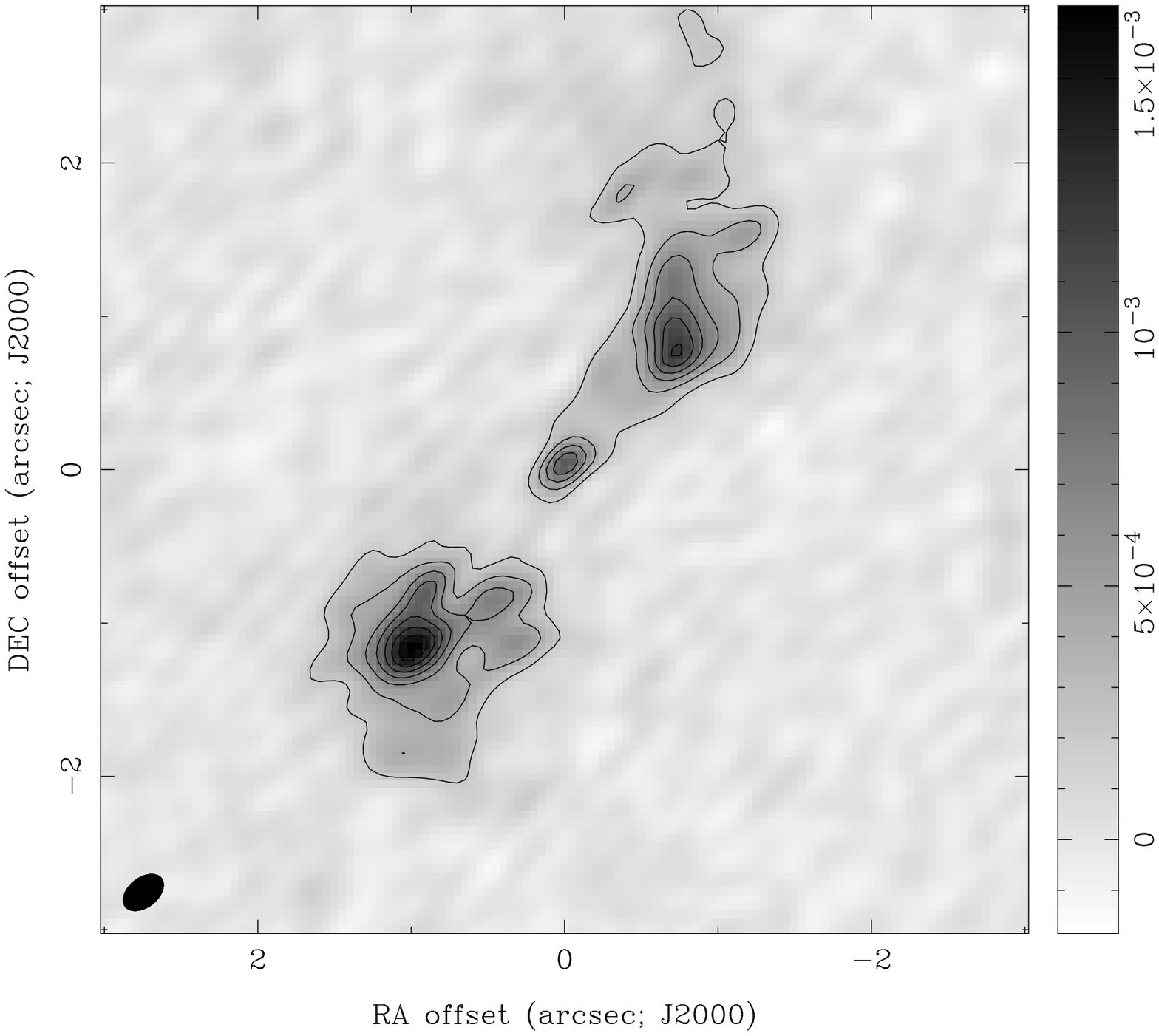}}
     \caption{Map of IRAS 22568+6141 at 8.4 GHz with contours at $50 \times (-3, 3, 6, 12, 15, 18, 21, 24, 27, 30) \; \mu$Jy~beam$^{-1}$.}
    \label{fig:22568map}
  \end{figure}

 {The nature of the continuum emission from IRAS~$22568+6141$ is more 
difficult to interpret from its spectrum, because of the presence 
of different components in the source, which were revealed by our 
high angular resolution map.}
{As displayed in Figure~\ref{fig:22568map}, this source includes a central compact component and two lobes in the north-west and south-east directions. The spectrum in Figure~\ref{fig:spectra} is integrated over the entire nebula, therefore new radio-continuum spectra of the three different components are needed to fully understand this object.
From the available data, we see that the flux densities from 3 GHz upward point to a steeply declining spectrum, matching again with the presence of a non-thermal process, while the points below 3 GHz appear almost flat.  Since optically thin emission below 1 GHz is not observed in any other of our targets and is indeed characteristic of extremely  evolved and diluted circumstellar envelopes of PNe, whose observational parameters do not match with $22568+6141$ (for example, the source is too dusty to be highly evolved), it is more plausible that flatness of the low-frequency spectrum results from it being the superposition of spectra with different behaviours (for instance, a central wind plus thermal free-free). }

\section{Variable sources}
\label{sec:variability}
\begin{figure*}
  \centering
    \subfloat[\label{17423time}]{\includegraphics[width=0.3\textwidth]{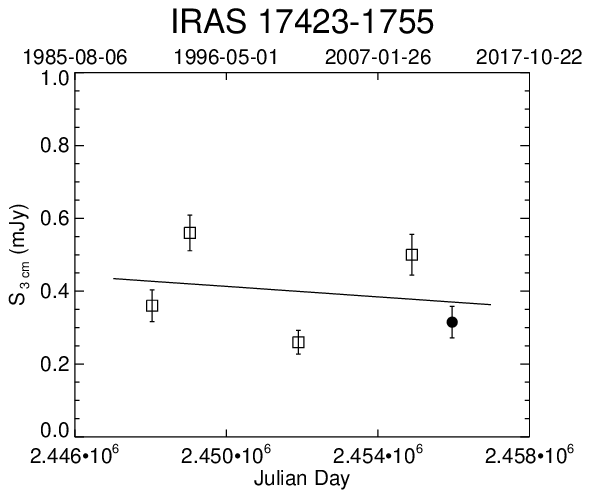}}\qquad
    \subfloat[\label{18062time}]{\includegraphics[width=0.3\textwidth]{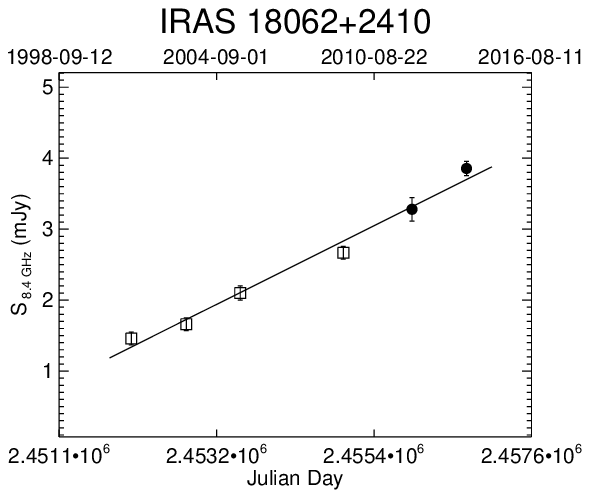}} \qquad
    \subfloat[\label{22568time}]{\includegraphics[width=0.3\textwidth]{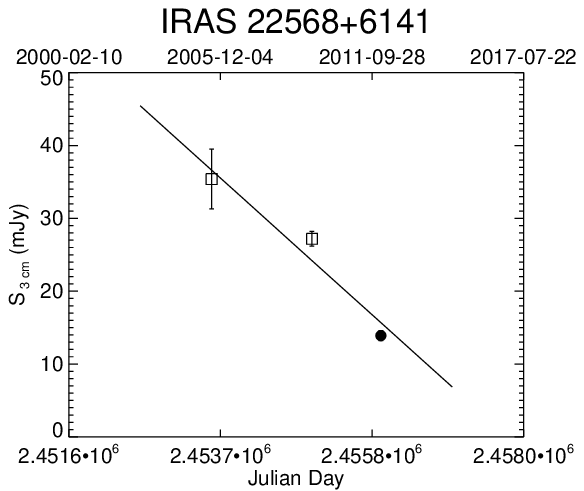}}
     \caption{Radio light curves of targets with evident flux-density variations. Open circles are measurements at 8.4 GHz (old VLA), while solid circles are at 9.0 GHz (new JVLA).}
    \label{fig:lightcurves}
  \end{figure*}

We summarise in Table~\ref{tab:epochs}, the results of our monitoring programmes and list the flux densities measured in those nebulae that display variations. In Figure~\ref{fig:lightcurves}, those variations are plotted for three sources with very different time patterns. Besides these targeted campaigns from 2001, 2009, and 2012, the light curves have been built using data from other runs presented by \citet{cerrigone11_radio}.

In \citet{cerrigone08}, we speculated that  IRAS $17423-1755$ has a periodic time pattern. Our new measurement confirms the variability but is inconclusive with respect to the time pattern. Observations with a much finer time baseline seem necessary to assess whether the variability is periodic or erratic. Since the flux calibration at this frequency is not very sensitive to atmospheric conditions, it is unlikely that the measured variations can be just due to observational errors, but rather point to an intrinsic instability. 

IRAS $18062+2410$ is the only target where the flux density at 8--9 GHz has been increasing systematically over the years. There is no doubt that the ionised mass in the nebula around this star is rapidly growing. Another target that {seems to display an increase} in flux density is $18371-3159$. For this source, we have two measurements 8 years apart at 8--9 GHz. Although this result should be confirmed with further measurements at the same frequency,  this source seems to be comparable to $18062+2410$.

\citet{ofek11} list IRAS $22568+6141$ among the variable sources detected in the NVSS survey \citep{condon88} with flux variations larger than $4\sigma$. The authors report that this nebula was observed three times over a period of about 35 days around 1996 Aug 30, displaying an average flux density at 1.4 GHz of $11.5 \pm 0.7$ mJy (including a 5\% error on the flux scale). The source was also observed with the VLA on 2002 Jul 25 at 4.8 GHz within the RMS survey \citep{urquhart09}, with its north-west and south-east lobes displaying  flux densities of $15 \pm 1$ and $14 \pm 1$ mJy, respectively. Finally, \citet{pazderska09} report a measurement of $31.4 \pm 4.1$ mJy at 30 GHz with the {OCRA-p} single-dish telescope, performed between December 2005 and May 2007.

At 1.4 GHz, the source shows an increase of flux density between 1996 and 2012. {Such an increase would be} compatible with an expanding ionisation front. In this case, the flux density is directly related to the size of the emitting surface and the variation  indicates that this is growing. The interpretation of the measurements at 4.8 and 30 GHz is more complicated and somehow challenges our finding of  a steep spectral index. 
The measurement at 4.8 GHz from 2002 gives a total flux density of $\sim$30 mJy. If we combine this information with our measurement at 8.4 GHz from 2005 and the OCRA-p value from $\sim$2006, the spectrum between 4 and 30 GHz seems to be flat within errors and stable over time. 

Such a stability over time is in contradiction with the source displaying a striking decrease in flux density between 2005 and 2012 and also with the variability in the NVSS. This strengthens the need for coeval multi-frequency data sets {at high angular resolution} to properly study this object and stresses the rapidity of the phenomena occurring in its environment.
At 8--9 GHz, the nebula is certainly fading rapidly  and its behaviour can be compared to what found in the Stingray Nebula. 

\begin{figure}
  \centering
    {\includegraphics[width=0.35\textwidth]{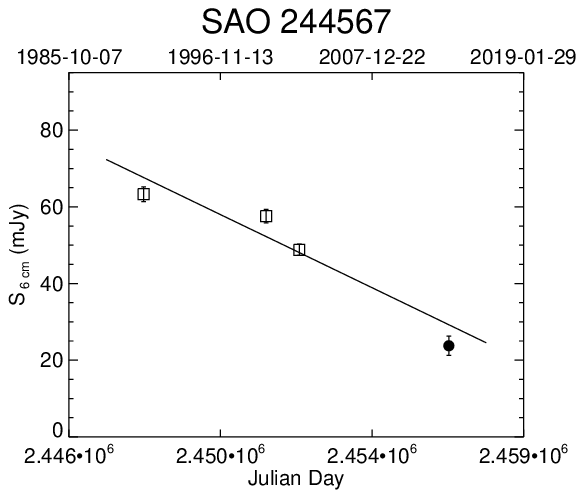}}
     \caption{Radio light curve of SAO 244567 at 8.6 GHz. The point from 2013 (solid circle) is a measurement at 19 GHz scaled down to 8.6 GHz, assuming a spectral index of $-0.1$.}
    \label{fig:saolightcurve}
  \end{figure}

As already mentioned, \citet{umana08} noticed that the radio emission from the Stingray was fading. We took the data from \citet{umana08} and complemented these with the new measurement set found in the web archive of the Australia Telescope Compact Array at 19 GHz. To compare it with the previous measurements of the Stingray, we  scaled this new point to 8.6 GHz,  assuming a spectral index of $-0.1$. 
The resulting light curve is plotted in Figure~\ref{fig:saolightcurve}, where we can see how rapidly the Stingray Nebula is fading and compare it to $22568+6141$ (Figure~\ref{22568time}). 

 IRAS $22568+6141$ lacks the great amount of spectral and photometric data available in the literature for the Stingray, but it appears as a second candidate to the category of {transition objects} that are actually fading away fast in the radio, pointing to an episodic ionisation event in their recent past.

\begin{table*} \centering
\begin{tabular}{lccc}
\hline
Target 			& 2001 			& 2009 			& 2012  \\
IRAS ID 			&   				& {S$_\nu$ (mJy)} 	&  \\
\hline
$06556+1623$ 		& 0.55$\pm$0.03 	& 0.48$\pm$0.04 	& 0.47$\pm$0.04  \\
JD 				& 2452071		& 2454988 		& 2455954 \\ \noalign{\smallskip}
$17423-1755$ 		&  0.26$\pm$0.03 	& 0.50$\pm$0.05 	& 0.32$\pm$0.04 \\
JD 				& 2452069 		& 2455006 		& 2456049 \\ \noalign{\smallskip}
$17516-2525$ 		&	-- 			&  0.32$\pm$0.06 	& 0.44$\pm$0.04 \\
JD 				& -- 				& 2455002 		& 2456049 \\ \noalign{\smallskip}
$18062+2410$ 		&  1.46$\pm$0.05 	& 2.67$\pm$0.09 	&  3.28$\pm$0.17 \\
 JD 				& 2452069 		& 2455006 		& 2455957 \\ \noalign{\smallskip}
$18371-3159$ 		&  0.62$\pm$0.03 	& -- 				&    0.67$\pm$0.06 \\
JD 				& 2452069 		& -- 				& 2456049 \\ \noalign{\smallskip}
$18442-1144$ 		& 19.2$\pm$0.6 	& 20.7$\pm$0.6 	& 21$\pm$1.0  \\
JD 				& 2452069 		& 2455006 		& 2455957 \\ \noalign{\smallskip}
$20462+3416$ 		&  0.42$\pm$0.03 	& 0.43$\pm$0.04 	& 0.47$\pm$0.03 \\ 
JD 				& 2452069 		& 2454992 		& 2455957 \\ \noalign{\smallskip}
$22568+6141$ 		& 32.7$\pm$0.98$^a$ & 21.7$\pm$0.7 	& 13.4$\pm$0.7 \\
$22568+6141$ NW 	& --$^b$ 			& 11.2$\pm$0.3 	&  5.7$\pm$0.3 \\		
$22568+6141$ SE 	&  --$^b$ 			& 9.3$\pm$0.3 		& 7.9$\pm$0.4 \\
JD 				& 2453580+2453585 & 2454992 		& 2455957 \\ 
\hline
$^a$Measurement from 2005.\\
$^b$The angular resolution was too coarse to disentangle the emission from the lobes.
\end{tabular}
\caption{Flux densities in X band (8.4/9.0 GHz) of the targets observed at different epochs with the Julian Day of the observation.}
\label{tab:epochs}
\end{table*}

\section{Individual sources}
In the following we will address four of our targets for which data spanning several different epochs are available. Three of these have also been imaged at high angular resolution.

\subsection{IRAS 18442-1144}
Among the targets in our study, IRAS~$18442-1144$ {(I18442)} is one for which we have enough measurements at different epochs to claim not only that flux density is stable over time within errors, but that its optically thin emission does not seem to be following the expected (almost flat) dependency on frequency typical of thermal emission.
By fitting the continuum points and applying statistical propagation of errors, we derive a spectral index of $-0.40~\pm~0.09$.  

\begin{figure*}
  \centering
    \subfloat[\label{bandax}]{\includegraphics[height=5cm]{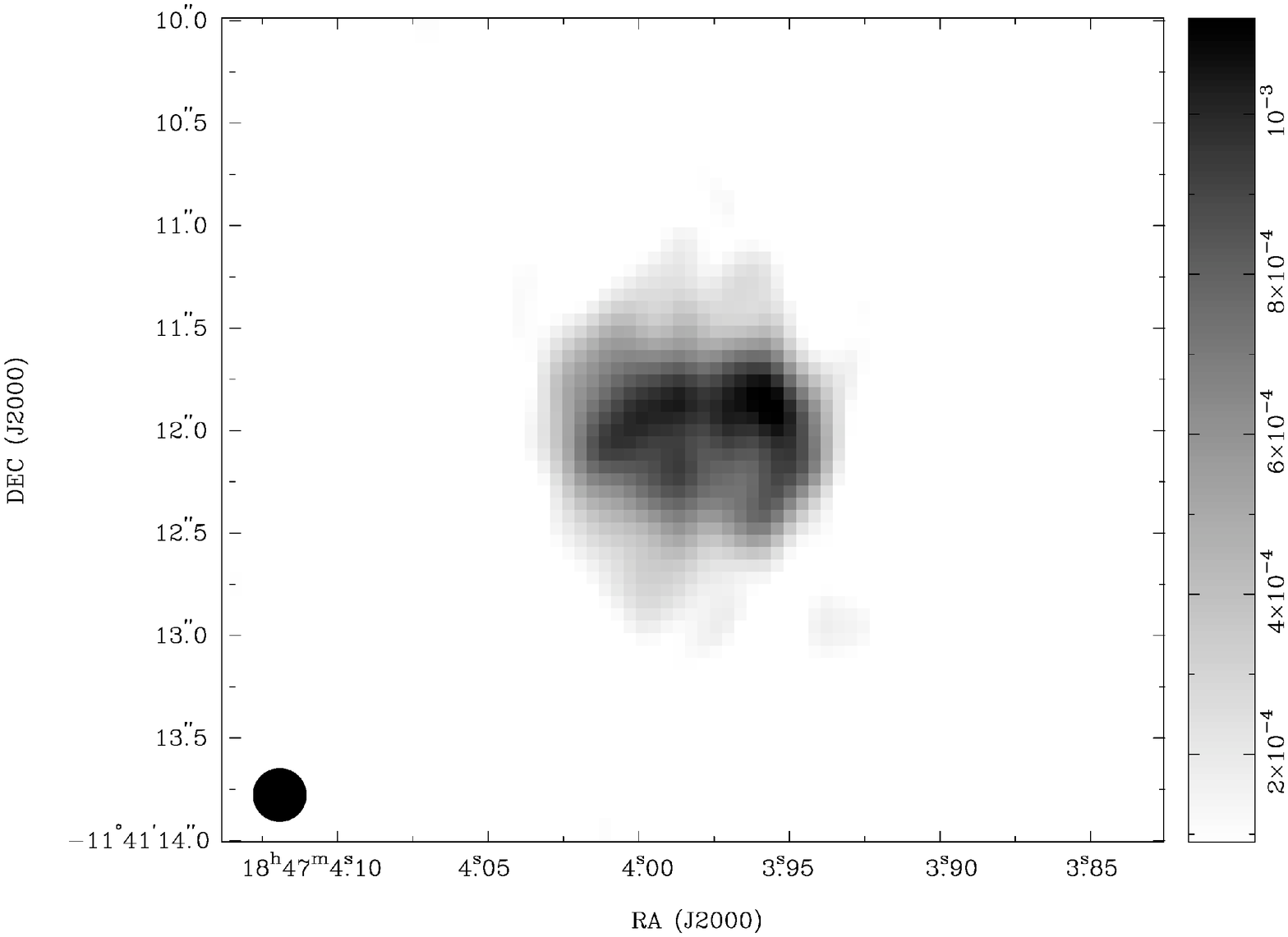}}\qquad \qquad
    \subfloat[\label{bandak}]{\includegraphics[height=5cm]{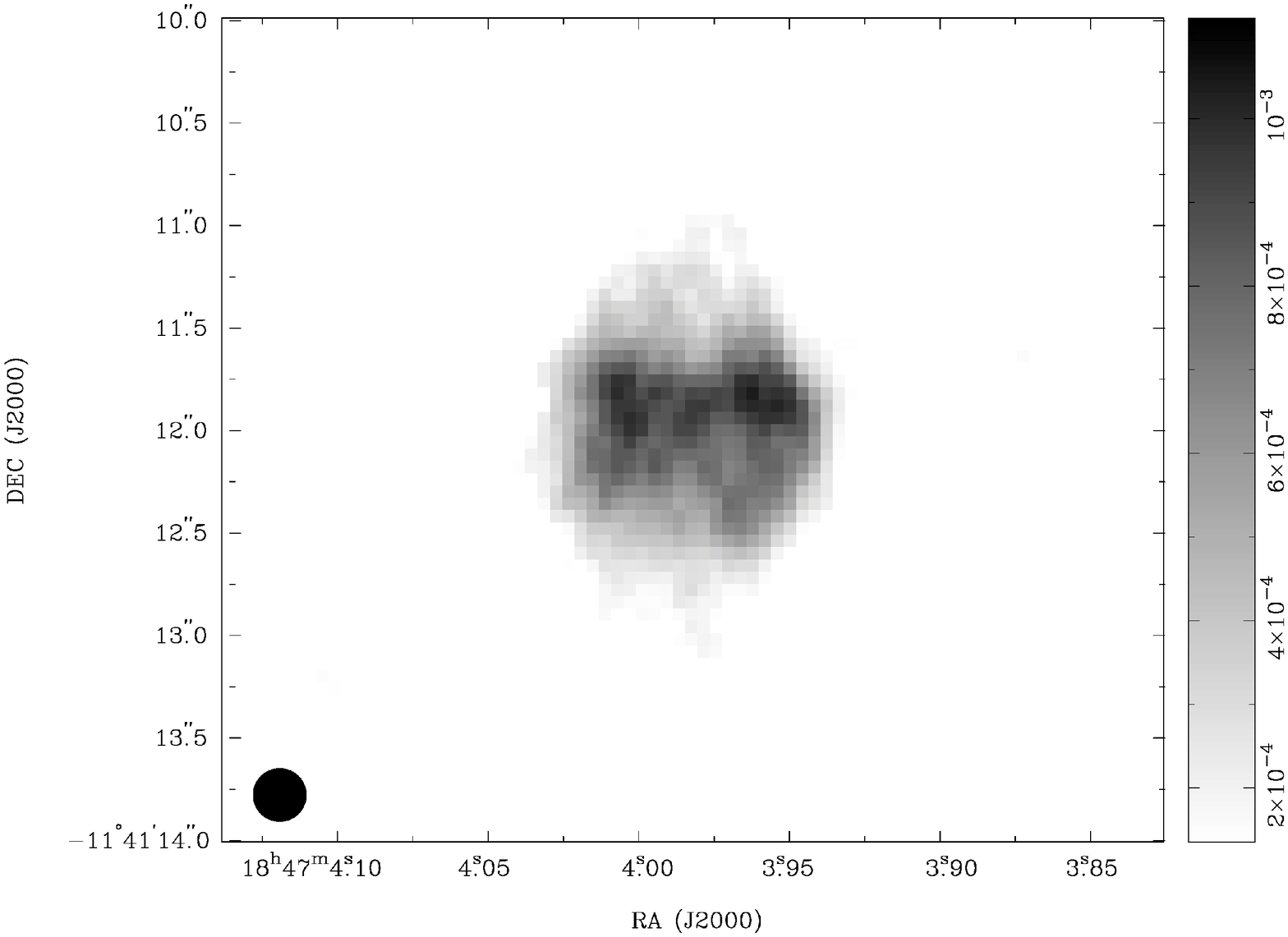}} 
    
    \subfloat[\label{spind}]{\includegraphics[height=5cm]{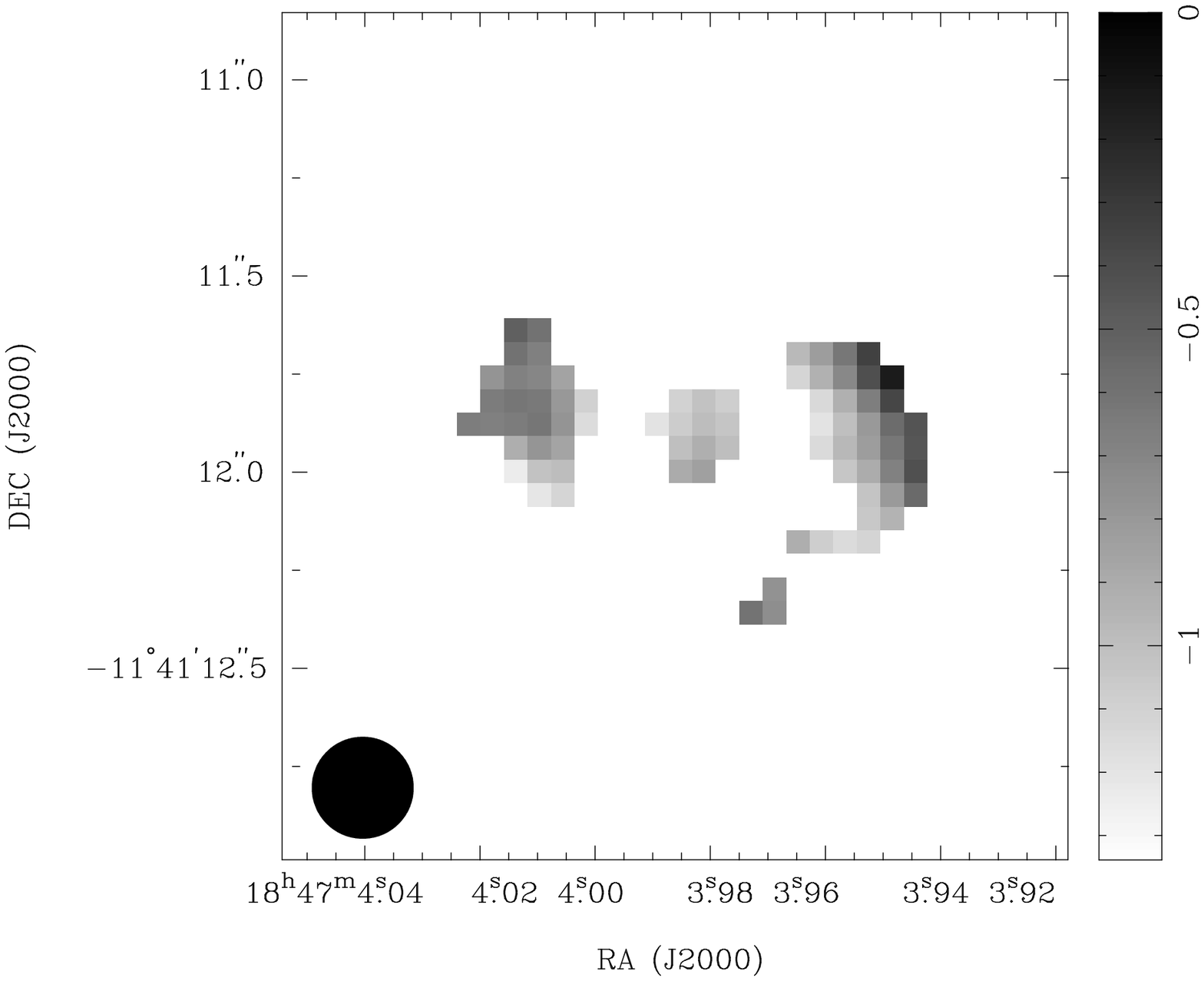}} 
     
    \caption{Maps of IRAS 18442-1144 at 8.4 GHz (Figure~\ref{bandax}) and 22.4 GHz (Figure~\ref{bandak}) and the spectral-index map between these two frequencies (Figure~\ref{spind}), constructed after cutting each of the previous maps at $5\sigma$.} 
    \label{fig:i18442maps}
  \end{figure*}

In \citet{cerrigone08}, we showed our map of I18442 at 8.4 GHz. Here, we will compare that map to another that we obtained by reprocessing our VLA observations at 22.4 GHz performed on 2004 October 22, when the array was in A configuration. The data set was calibrated in CASA and the application of multi-scale cleaning and tapering resulted in a much better map than what was achievable at the time the observation was carried out. This is due to the fact that in A array at 22.4 GHz, the VLA resolved this source and multi-scale cleaning is able to better recover extended emission. In Figure~\ref{fig:i18442maps}, we compare the two maps at 8.4 and 22.4 GHz, which were created with the same cell size and restoring beam, for a proper pixel-to-pixel comparison.

The emission peaks at both frequencies are found in an open ring that extends approximately from south-west to north-east. The ring is clumpy and weaker emission extends around it parallel to its axis, following the typical  shape observed in many post-AGB nebulae with lobes.

Since we have two maps at two different frequencies, we want to investigate if the spectral index changes within the nebula. We created our maps so that such a comparison can be done pixel by pixel and display  the spectral-index map in Figure~\ref{fig:i18442maps}. 

The spectral-index map was created after cutting each of the two VLA maps at the $5\sigma$ level, to include only the brighter components. The final map shows three different regions
of emission, one at the centre and  two located east and west from it. The eastern and western structures appear as clumps in the circumstellar ring, while the central component can be interpreted as located in the direction of the central star.

We boxed  each of these three components with a polygon, then calculated the average spectral index within them and the relative standard deviation, obtaining $-1.0\pm0.1$, $-0.8\pm0.3$, and $-0.8\pm0.2$ for the central, western, and eastern components, respectively.

 Interestingly, the central component has a smaller standard deviation and the smallest value of spectral index, while the opposite is true for the two circumstellar components  (they have larger spectral index and standard deviation). These values strengthen the visual appearance in which the circumstellar structures span a wider range of spectral-index values, with a trend for higher values toward their outer edges and smaller values toward the central component, which instead appears as a smooth grey area in the map. 
 
 Our interpretation is therefore that the non-thermal emission peaks on the central star (or more generally in the inner nebula) and then turns more and more to thermal toward the outer nebula. This differentiation explains why the overall spectral index is around $-0.4$, when also including regions at lower brightness (farther away from the centre).

\subsection{IRAS 20462+3416}
IRAS $20462+3416$ displays a critical frequency around 2--3 GHz and its flux density falls rapidly with increasing frequency. 

In February 2003, we measured for this source a flux density at 22.4 GHz that does not match with the values observed at the other frequencies. 
We reprocessed the 2003 data set and compared the result of our new reduction with our previous result and with that calculated from the data processed through the VLA pipeline. The flux densities for the three sets match within errors. 

Since observations at 22.4 GHz suffer particularly from atmospheric opacity and the flux calibrator (3C~286) is about $87^\circ$ away from this source, we conclude that our 2003 measurement at 22.4 GHz cannot be considered reliable, likely because of the atmospheric conditions. This stresses the care that must be exercised in interpreting continuum slopes when only one measurement is available in a frequency band, especially at high frequencies. 

The fit to the continuum of this target in its optically-thin range in Figure~\ref{20462spec} is obtained neglecting our {upper limit} at 45 GHz. By fitting the continuum points between 4 and 20 GHz and applying statistical propagation of errors, we derive a spectral index of $-0.37~\pm~0.02$. 

The line thus fitted to the data {slightly overestimates our upper limit} at 45 GHz.  Albeit  keeping in mind the care mentioned above, we consider that our non-detection at high frequency of IRAS $20462+3416$ is a confirmation  that this source has a steep continuum spectrum indicating non-thermal emission and deserves further attention.


\subsection{IRAS 18062+2410}
IRAS $18062+2410$ is a hot post-AGB star with clear signs of an ionised circumstellar region. 
Both its optical recombination lines (H$\alpha$ and H$\beta$) and its radio flux display variability over a few years. \citet{gledhill15}  observed it with NIFS@Gemini and found that its compact and unresolved ionised region is surrounded by an axisymmetric molecular envelope. 

In \citet{cerrigone11_radio}, we showed that the radio flux density around 8 GHz has steadily increased since 2002, a behaviour that is comparable to what observed in the well-known post-AGB object AFGL 618. The O-rich $18062+2410$ could be a scaled-down (i.e., at lower mass) version of the C-rich AFGL 618.

\begin{figure*}
  \centering
    \subfloat[\label{18062vla}]{\includegraphics[height=5cm]{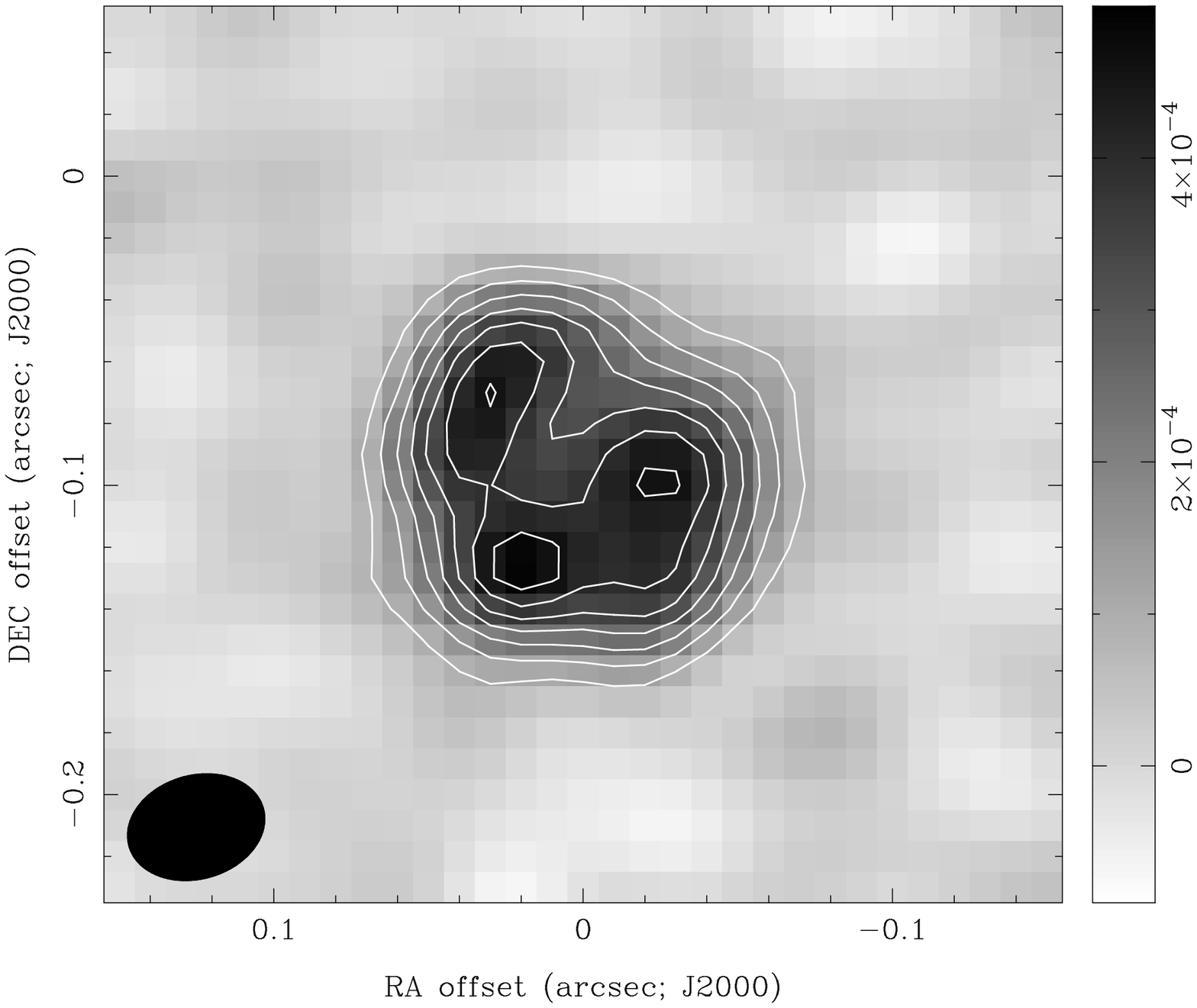}}\qquad \qquad
    \subfloat[\label{18062vlamodel}]{\includegraphics[height=5cm]{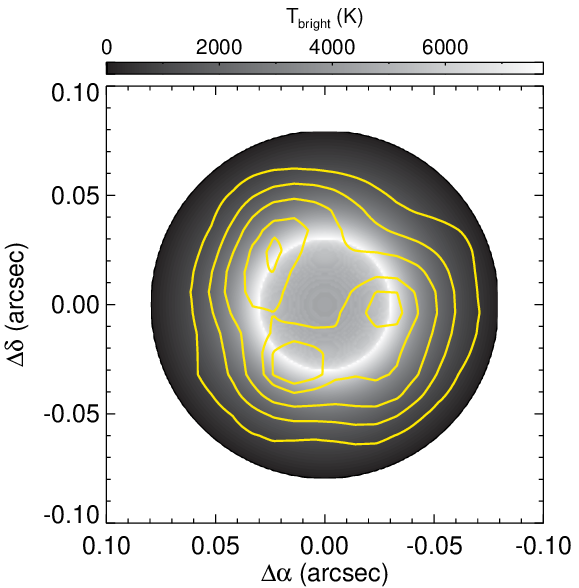}} 
     \caption{Map of IRAS 18062+2410 at 45 GHz ($\theta_{syn}\sim 0.04''$) with overlayed contours at $30\times(3, 4, 5, 6, 7, 8, 9, 10) \; \mu$Jy~beam$^{-1}$ (Figure~\ref{18062vla}) and with contours overlayed on an emission model map (Figure~\ref{18062vlamodel}).}
    \label{fig:i18062maps}
  \end{figure*}

Following our previous modelling, we observed $18062+2410$ with the JVLA at 7 mm in 2014 in the most extended array configuration, obtaining a synthesised beam of $\sim$0.04$''$. The small beam allowed us to resolve its ionised region.  As can be seen in Figure~\ref{18062vla}, this is a roughly circular shell containing three main clumps with an overall diameter of about 0.16$''$ ($1.5\times10^{16}$ cm at the estimated distance of 6.4 kpc). For a comparison, the ionised region in AFGL 618 has a diameter of  {about 0.5$''$ ($6.8\times10^{15}$ cm} at a distance of 0.9 kpc). 

Given the size constraints of our new high-resolution data, we performed new modelling of the source {as described in \citet{cerrigone11_radio}, but with radii derived from the new maps: inner radius of 0.03$''$ and outer of 0.08$''$. As previously done, the density profile was set to follow a power law of the type r$^{-2}$ and matching by eye both its spectrum and its brightness map}. The results of the modelling are shown in Figure \ref{18062vlamodel}. Our model returns a value of the ionised mass of $3.2\times10^{-4}$~M$_\odot$ and a density at the inner radius of $7.7\times10^5$~cm$^{-3}$.  {We estimate an uncertainty of about 15\% on the density at the inner radius and the total mass, by calculating values of the density that would double the ${\chi}^2$ of the fit to the spectrum}.

 We can compare our radio map to the near-infrared images obtained by \citet{gledhill15}. They detected two peaks of H$_2$ emission lines $\sim$0.25$''$ off-set from our radio position, roughly along the south-west to north-east direction. These appear as clumpy blobs of emission at either side of the radio ring that we detected with the JVLA. 
  
  Though striking at first, the morphological difference between the near-IR and radio images needs to be weighted by the extreme compactness of the ionised region and consequent limited resolving power. 
The radio ring in our map looks like a clumpy spherical shell projected on the plane of the sky. Besides its being clumpy, we can  identify two halves in it. One region with two clumps south-west of the centre and another region with only one peak north-east of it. 

 Although somewhat speculative with the current maps, this leads us to conclude that the regions seen in the H$_2$ lines and in the radio are roughly aligned along the same direction and are what remains of the circumstellar envelope after the removal of material in the direction perpendicular to them, occurred at an earlier evolutionary stage. In this scenario, the current emitting regions would trace what used to be the circumstellar waist in the early post-AGB, when an outflow probably pierced and swept away whatever was located along the north-west/south-east direction.   {If this were confirmed by further observations, giving better knowledge of the geometry of the radio structure, it would be necessary to correct our modelling with an emitting torus rather than a circumstellar shell. }

\subsection{IRAS 22568+6141} \label{sec:i22568}
Our JVLA data confirm that IRAS~$22568+6141$ {(I22568)} is steadily decreasing in flux density around 8 GHz. 

The behaviour of  {I22568} is similar to what described by \citet{zijlstra08} for the planetary nebula NGC~7027, with the flux density increasing at low frequency and decreasing at the higher ones. Yet, the time scale in  {I22568} is too fast and the effect too dramatic for the phenomenon to be the same (the flux variations in NGC~7027 are fractions of 1\% over decades). As mentioned already, we believe that the radio spectrum in $22568+6141$ points to the presence of a non-thermal emission mechanism.


\citet{sanchez-contreras12} detected CO emission in the J=$1-0$ line toward I22568. Interestingly, they also detected continuum emission at 115 GHz with a flux density of $9.8\pm1.4$ mJy. They carried out their measurements between March and April 2003 with CARMA. The largest angular scale of their observation was about 38$''$ ($5.4 \mathrm{k}\lambda$ in the $uv$ plane), which assures that the interferometer was not filtering out extended emission, since the nebula is about 8$''$ across.

\begin{figure}
  \centering
    {\includegraphics[width=0.4\textwidth]{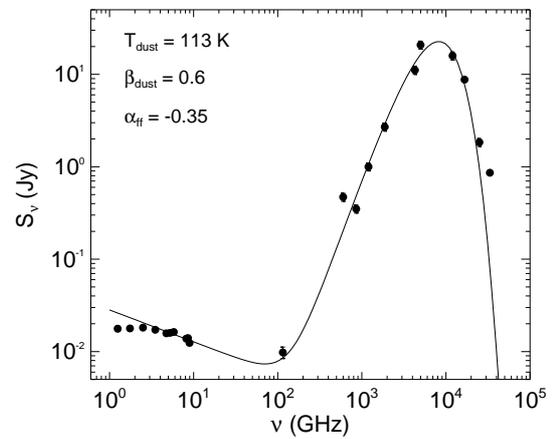}}
     \caption{SED of IRAS 22568+6141. The solid line is the sum of the fit to the thermal dust emission and that to the {high-frequency} radio measurements.}
    \label{fig:22568sed}
  \end{figure}

 At 115 GHz, the emission is likely a combination of radiation from dust and ionised gas. To investigate this, we plotted our radio points along with that at 115 GHz and those from the IRAS \citep{iras} and AKARI \citep{akari_mid} catalogues, at 12, 25, and 60 $\mu$m and at  9 and 18 $\mu$m, respectively. To fill the far-IR gap, we searched the Herschel archive and found that the Galactic location of IRAS~$22568+6141$ falls within the area covered by the Hi-Gal survey \citep{higal}. We then retrieved pipeline-processed Herschel images at 70, 160, 250, 350, and 500 $\mu$m, performing aperture photometry on them and obtaining 11.1, 2.7, 1.0, 0.35, and 0.47 Jy, respectively; we estimated errors of $\sim$10\% at all wavelengths, accounting for both an absolute error of $\sim$5\% \citep{molinari16}  and the image noise. At 500 $\mu$m,  the intense variations in the field around the target make it difficult  to properly estimate the background emission and clearly cause an overestimate of the flux density.  The results are plotted in Figure~\ref{fig:22568sed}.

The {spectral energy distribution of this source can be reproduced} combining the emission from the dust  and ionised regions. To this aim, several approaches can be followed, like using a code that includes both dust and ionised gas, or a combination of two different codes for the two components, or finally simply an empirical fit to the observational points. The use of a code for the dust implies the need to know \textit{a priori} the optical properties of the grains. In evolved stars, such properties are not very clear and \textit{ad hoc} corrections to standard optical constants are often necessary \citep{ladjal10, umana_planck}. Since a detailed knowledge of the dust properties of this source is not our scope,  we modelled our target by fitting a modified Planck curve (a Planck curve multiplied by the factor  $\nu^\beta$, to account for the dust emissivity) to the IR data points, excluding that at 9 $\mu$m, which is probably affected by stellar contributions and/or strong spectral features. This fit returned values of 113 K for the average dust temperature and 0.6 for the dust emissivity index. Such small values of the emissivity index are typical of evolved stars \citep{guertler96}.

Once we fitted the IR data points, we fitted the radio points. 
After summing the fits to the radio and IR points, we find that the measurement at 115 GHz is well matched, which is surprising, if we consider that  it was performed in 2003. At 115 GHz, flux contributions from both dust and ionised gas can be relevant.  The dust emission is not expected to vary over decades. From our fit to the infrared data, we derive that at 115 GHz the dust contributes  $\sim$3~mJy to the total flux density, therefore most of the emission at this frequency is due to ionised gas. As the ionised component has been fading since 2005, {one would  expect that the fit of the 2012 high-frequency radio points summed with the dust emission should underestimate the measurement at 115 GHz performed in 2003. 
We interpret this fact as the ionised emission being due to both a thermal and a non-thermal component. Since the non-thermal emission has a steep spectrum, at 115 GHz the thermal contribution should be dominant. Therefore, if the flux density at 115 GHz indeed does not decrease with time, we can assume that the thermal component is constant, while the variability around 10 GHz is due to decreasing non-thermal emission from a past episodic event.
 Such a scenario could be easily  confirmed or ruled out by a new measurement at 115 GHz. }
 
 The {millimetric measurement} dates back to 2003, therefore we compare its value to the flux density obtained at 4.8 GHz in 2002 by \citet{urquhart09}, which amounts to $29 \pm 1$ mJy. A purely thermal spectrum between 4.8 and 115 GHz would imply a flux density at the high frequency of about 21 mJy, more than the double of what measured  with CARMA only one year later.

\begin{figure}
  \centering
    {\includegraphics[width=0.4\textwidth]{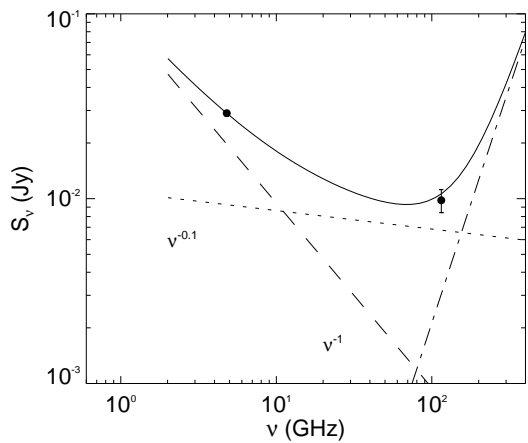}} 
     \caption{The SED of 22568 interpreted as the sum of emission from dust (dot-dashed), thermal free-free (dotted), and a non-thermal component (dashed) around 2002--2003.}
    \label{fig:22568specthen}
  \end{figure}

As mentioned already, we constrain the thermal emission by assuming it accounts for all the emission at 115 GH that is not due to dust ($\sim$6.8 mJy). We then calculate the thermal contribution at 4.8 GHz and finally add up a non-thermal spectrum to match the total flux density at this frequency. With the assumption of a negligible non-thermal contribution at 115 GHz, the non-thermal spectral index cannot be larger than $-1$, not to substantially overestimate the millimetric measurement (Figure~\ref{fig:22568specthen}).

\begin{figure}
  \centering
    {\includegraphics[width=0.4\textwidth]{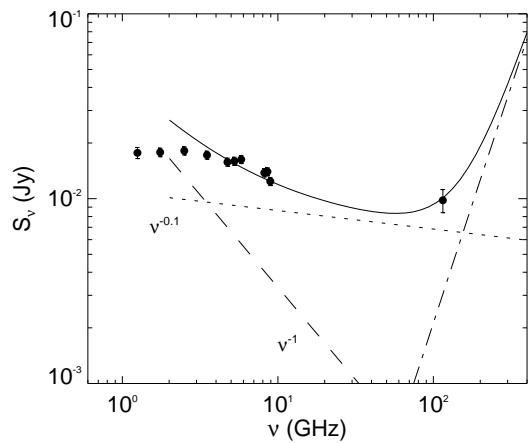}}    
     \caption{The SED of 22568 interpreted as the sum of emission from dust (dot-dashed), thermal free-free (dotted), and a non-thermal component (dashed) in 2012.}
    \label{fig:22568specnow}
  \end{figure}

As an example, we show in Figure~\ref{fig:22568specnow} how the sum of a thermal and a non-thermal components could match the observations, assuming that {only} the dimming of the non-thermal contribution  is  responsible for the flux decrease. New multi-frequency observations in the next years will possibly confirm this scenario until the emission levels off to purely thermal. A relevant point will also be investigating the spectral and time behaviour of the three different sources of emission that we detected in this object at high angular resolution.

\section{Summary}
We have investigated the radio variability of a sample of { stars transiting between the post-AGB and the PN phases}  that we have observed over several years with the Very Large Array.
Our results show that in a few cases variability is detected and it displays both decreasing and increasing emission.

A systematically increasing pattern is detected only in IRAS $18062+2410$, whose morphology we have resolved at 45 GHz, finding that the nebula has clumps of emission approximately matching the lobe-like distribution observed by other authors through infrared recombination
lines, although the radio-emitting region is much more compact than that detected in the infrared range.

Spectral indices smaller than the typical thermal value of $-0.1$ for ionised shells have been detected in IRAS $18442-1144$, IRAS~$20462+3416$, and IRAS~$22568+6141$. 
Two of these targets have been mapped at high angular resolution and display two very different morphologies. IRAS~$18442-1144$ has a clumpy equatorial waist and the ionised emission arises from this waist and the nebula along its axis. A comparison of the radio maps at 8.4 and 22.4 GHz of this object points to the spectral index being flatter toward the outer edge of the nebula and steeper in the inner region. This target does not display radio variability and its spectral index is not particularly large in absolute value, which may be interpreted as the remaining effect of a non-thermal component that is about to disappear.

When mapped at high angular resolution, IRAS~$22568+6141$ displays three emitting structures: a central compact source and two blobs of emission north-west and south-east of it. Most of the radio emission comes from the two off-set blobs. Thanks to the combination of ours and literature data, we estimate that the decreasing flux density in this source is probably due to a non-thermal radio component that is dimming quickly on top of a thermal continuum. 

If we link the variability to the existence of a non-thermal component, IRAS~$22568+6141$ may be seen as still closer to the episode that caused the formation of this component, while IRAS~$18442-1144$ may possibly be seen when this is about to end. In this speculative scenario, one can wonder whether IRAS~$18062+2410$ is instead being observed before the event leading to shock formation  has occurred.
\citet{qiao16} detected the OH maser line at 1720 MHz in the young PN  IRAS~$16333-4807$, with a Zeeman splitting pointing to a magnetic field with strength of about 2--10 mG. Since the line is unusual in evolved stars and its velocity pattern does not match with that of the two other OH transitions in the same nebula, the authors conclude that the feature at 1720 MHz is likely to be caused by an equatorial ejection propagating in a magnetised environment. This ejection possibly occurred close to the onset of the ionisation. Similar events may also link the radio variability and non-thermal components found in this work.

The detection of radio variability and its connection with non-thermal components is a topic that certainly deserves further investigation in these and more sources in the next years, thanks to the widely improved performance of radio telescopes. Moreover, the new radio facilities coming up in the next years such as ASKAP and SKA will be able to detect more sources in these peculiar evolutionary phases, thanks to their sensitivities and abilities to rapidly map wide areas of the sky \citep{umana_ska}. Even before these new telescopes become fully operational, exploratory studies like the SCORPIO survey \citep{scorpio} can be exploited to look for more targets in the attempt to broaden the limited sample currently available.

\section*{Acknowledgments}
We thank an anonymous referee for comments that improved this paper. The Australia Telescope Compact Array is part of the Australia Telescope National Facility which is funded by the Australian Government for operation as a National Facility managed by CSIRO.
This paper includes archived data obtained through the Australia Telescope Online Archive (http://atoa.atnf.csiro.au).
This research has made use of the SIMBAD database, operated at CDS, Strasbourg, France.
This research has made use of the VizieR catalogue access tool, CDS, Strasbourg, France

\bibliography{mybib}{}
\bibliographystyle{mnras}
\bsp

\label{lastpage}

\end{document}